\documentclass[prd,twocolumn]{revtex4}

\usepackage{psfrag}
\usepackage{graphicx}
\usepackage{graphics}
\usepackage{bm}
\usepackage{color}
\usepackage{verbatim,color,ulem}

\newcommand{\be}{\begin{equation}}
\newcommand{\ee}{\end{equation}}
\newcommand{\ba}{\begin{eqnarray}}
\newcommand{\ea}{\end{eqnarray}}

\newcommand{\dcom}[1]{}
\newcommand{\dnote}[1]{}

\newcommand{\gsim}{\raise.3ex\hbox{$>$\kern-.75em\lower1ex\hbox{$\sim$}}}
\newcommand{\lsim}{\raise.3ex\hbox{$<$\kern-.75em\lower1ex\hbox{$\sim$}}}

\begin{document}

\renewcommand{\thefootnote}{\fnsymbol{footnote}}


\renewcommand{\thefootnote}{\arabic{footnote}}
\setcounter{footnote}{0} \typeout{--- Main Text Start ---}

\title{Nonequilibrium Damping of Collective Motion of Homogeneous Cold Fermi Condensates with
Feshbach Resonances}
\author{ Chi-Yong  Lin and Da-Shin  Lee}\affiliation{
Department of Physics, National Dong Hwa University, Hua-Lien,
Taiwan 974, R.O.C.  } \author{Ray\ J.\ Rivers}
\affiliation{ Blackett Laboratory, Imperial College\\
London SW7 2BZ, U.K.}

\date{\today}
\begin{abstract}
Collisionless damping of a condensate of cold Fermi atoms, whose
scattering is controlled by a Feshbach resonance, is explored
throughout the BCS and BEC regimes when small perturbations on its
phase and amplitude modes are  turned on to drive the system
slightly out of equilibrium. Using a one-loop effective action, we
first recreate the known result that for a broad resonance the
amplitude of the condensate decays as $t^{-1/2}$ at late times in
the BCS regime whereas it decays as $t^{-3/2}$ in the BEC regime. We
then examine the case of an idealized narrow resonance, and find
that this collective mode decays as $t^{-3/2}$ throughout both the
BCS and BEC regimes. Although this seems to contradict earlier
results that damping is identical for both broad and narrow
resonances, the breakdown of the narrow resonance limit restores this universal behaviour. More measureably, the phase
perturbation may give a shift on the saturated value to which the
collective amplitude mode decays, which vanishes only in the deep
BCS regime when the phase and amplitude modes are decoupled.

\end{abstract}


\maketitle

\section{Introduction}
We now have remarkable experimental control over cold alkali
atoms interacting through a Feshbach resonance. Manipulation of the binding energy through external magnetic fields enables us to evolve them continuously
from the weakly coupled BCS-like behavior of Cooper pairs to the strongly coupled Bose-Einstein
Condensation (BEC) of molecules \cite{exp_MolecularBEC}. The
transition is characterised by a crossover in which, most simply,
the $s$-wave scattering length $a_S$ diverges as it changes sign
\cite{exp_Crossover,chin}.

Recently, a considerable theoretical effort
has been expended on understanding how such a tunable superconductor/condensate responds to an initial
non-equilibrium change in its order parameter. The behavior of tunable gases can differ according as the Feshbach resonance is broad or narrow, the former essentially describing a one-channel system, the latter a two-channel system.
 Nonetheless, in~\cite{gur1}  it is argued that, for linear perturbations, the relaxation through damping of this collective
motion is the same for both broad and
narrow resonances.  What determines the nature of the damping is whether the chemical potential is positive (BCS regime) or negative (BEC regime). In the former case it is proposed that the perturbed amplitude of the condensate decays with elapsed time $t$ as $t^{-1/2}$, as happens for superconductors~\cite{vol}, and in the latter as $t^{-3/2}$.
Advanced detection techniques certainly permit experimental
time-resolved studies of the collective pairing mode~\cite{chin}
 with regard to such perturbations~\cite{gur1}. [There has also been work on non-linear perturbations ~\cite{yuz,bar1,bulgac}, but this goes beyond the analysis that we shall present here.]

Here we provide an extensive study to reexamine  the relaxational
dynamics of the amplitude mode of the fermionic condensate as a
result of this (Landau) damping throughout both BCS and BEC regimes
for both broad and narrow resonances, building on our earlier work~\cite{rivers,rivers1}
on the BEC/BCS crossover in the presence of a Feshbach resonance.
In~\cite{rivers,rivers1} we have shown that many condensate
properties at T = 0 can be derived easily from the derivative
expansion of the renormalized one-loop effective action. In
particular,   under long wavelength and small energy approximations
this action can be used to obtain
 a hydrodynamic description of the BEC-BCS crossover from which the
equation of state, intimately related to the speed of sound, can be
derived.  In this paper we particularly build on the work of
\cite{rivers}, creating an extended {\it full} one-loop
effective action  to explore the real-time evolution of the
collective mode.

We find that the situation is more subtle than as presented in~\cite{gur1}, with some differences in the relaxational dynamics of the
amplitude mode  for broad and narrow resonances in the BCS regime
with positive chemical potential.
 For broad resonances we recreate the results of~\cite{gur1} directly.
 However, for idealised narrow resonances we find a $t^{-3/2}$ decay
across both BCS and BEC regimes. Nonetheless, we anticipate that
 for more realistic resonances the long-time behaviour of ~\cite{gur1} will be
 recovered in the deep BCS regime, albeit with $t^{-3/2}$ transients. A new result is that, even under linear perturbations, the amplitude mode is
found to decay to a shifted value from its initial condition, due to
the phase-amplitude coupling, irrespective of whether the resonance is broad or narrow. Such a shift is expected for
non-linear perturbations but not for linear perturbations, when
typically the system relaxes to its initial equilibrium state. These
findings can be tested experimentally.

\section{Effective Actions}
We consider a condensate comprising a mixture of fermionic atoms and
molecular bosons, in which the fermions $\psi_{\sigma} (x)$, with
spin $\sigma = ( \uparrow , \downarrow )$, undergo self-interaction
through an s-wave BCS-type  term, and two fermions can be bound into
a molecular boson $\phi(x)$ through a Feshbach resonance. The
general 'two-channel' microscopic action is given by ($U>0$, $g$
fixed)
 \begin{eqnarray}
S &=& \int dt\,d^3x\bigg\{\sum_{\uparrow , \downarrow}
\psi^*_{\sigma} (x)\ \left[ i \
\partial_t + \frac{\nabla^2}{2m} + \mu \right] \ \psi_{\sigma} (x)
\nonumber \\
& +& U \ \psi^*_{\uparrow} (x)\  \psi^*_{\downarrow} (x) \
\psi_{\downarrow} (x) \ \psi_{\uparrow} (x)
\nonumber \\
   &+& \phi^{*}(x) \ \left[ i  \ \partial_t + \frac{\nabla^2}{2M} + 2 \mu -
\nu \right] \ \phi(x) \nonumber \\
&-& g \left[ \phi^{*}(x) \ \psi_{\downarrow} (x) \ \psi_{\uparrow}
(x) +  \phi(x) \ \psi^{*}_{\uparrow} (x) \ \psi^{*}_{\downarrow} (x)
\right]\bigg\} \label{Lin}
\end{eqnarray}
Thus, the bound bosons of Feshbach resonance (field $\phi$) have  twice the mass of
the fermions, $M=2m$,  and a tunable binding energy, $\nu$.

On introducing the auxiliary field $\Delta (x) = U\psi_{\downarrow}
(x)  \psi_{\uparrow} (x)$, a Hubbard-Stratonovich transformation
leads to an effective Lagrangian density
 ${S}$ quadratic in the Fermi fields. We then integrate them out \cite{rivers}  to write ${S}$ in the non-local form
  \begin{eqnarray}
 S_{NL} &=& -i\,Tr\ln {\cal G}^{-1} + \int dt~ d{\bf x}\bigg\{- \frac{1}{U}|\Delta |^2\nonumber
 \\
 &+& \phi^{*}(x) \ \left[ i  \ \partial_t + \frac{\nabla^2}{2M} + 2 \mu -
\nu \right] \ \phi(x)\bigg\},
 \end{eqnarray}
 in which
  ${\cal G}^{-1}$ is the
inverse Nambu Green function,
\begin{equation}
 {\cal G}^{-1} = \left( \begin{array}{cc}
        i \partial_t - \varepsilon         & \tilde{\Delta}(x) \\
                  \tilde{\Delta}^{*}(x) & i \partial_t +
                  \varepsilon
                  \end{array} \right)  \label{greenfun}
                  \end{equation}
 where
 \begin{equation}
 \tilde\Delta (x) = \Delta (x) - g\,\phi (x) \label{tildeDelta}
 \end{equation}
 that represents the two-component {\it combined } condensate (and
$\varepsilon = - \nabla^2/2m - \mu $).

The combined condensate amplitude and phase of $\tilde{\Delta}(x) =
|\tilde{\Delta}(x)| \ e^{i \theta_{\tilde{\Delta}}(x)}$ can be
determined from those of $\Delta(x)$ and $\phi(x)$, defined
respectively by $\Delta(x) = |\Delta(x)|\ e^{i \theta_{\Delta}(x)}$
and $\phi(x) = -|\phi(x)| \ e^{i \theta_{\phi}(x)}$.  The $U(1)$
invariance of the action under
$\theta_{\Delta}\rightarrow\theta_{\Delta} + \rm{const.}
 ,\,\theta_{\phi}\rightarrow\theta_{\phi}+\rm{const.}$
is spontaneously broken. Thus,
 $\delta\,S_{NL} = 0$ permits spacetime constant {\it gap} solutions $|\Delta
 (x)|=|\Delta_0|\neq 0$ and $|\phi (x)| = |\phi_0|\neq
 0$ (whereby $(|{\tilde\Delta} (x)|=|{\tilde\Delta}_0|\neq
 0)$.

We now  consider the fluctuations around the gap configurations
and simultaneously perturb in the derivatives of $\theta_\Delta$ and $\theta_\phi$,
{\it small} perturbations in the scalar condensate densities
\cite{aitchison} $\delta |\Delta|= |\Delta| - |\Delta_{0}|$ and
$\delta |\phi|= |\phi|  - |\phi_{0}|$ and their derivatives. To guarantee Galilean invariance we expand~\cite{aitchison,rivers}in $\Sigma = {\cal G}^{-1} - {\bar{\cal G}}^{-1}$, where ${\bar{\cal G}}^{-1}$ is obtained from ${\cal G}^{-1}$ by substituting $\tilde{\Delta}(x)$ with the Galilean scalar $|\tilde{\Delta}(x)|$.
We construct the condensate
effective
 action $S_{\rm eff}$ at second order in $\Sigma$.
 Diagrammatically, this amounts to taking account of fermionic
one-loop effects in the effective action.

$S_{\rm eff}$ comprises the integral of a local density together with non-local fermionic cut contributions. Later we shall restrict ourselves to homogeneous collective modes.
For their dynamics we only need the quadratic part of the local density, while retaining the full cut contributions.
The relevant $S^{(2)}_{\rm eff}$ then takes the form
\begin{eqnarray}
 && S^{(2)}_{\rm eff} = \int d^4x\bigg\{
- \frac{1}{8m} \rho^F_0(\nabla \theta_{\tilde{\Delta}} )^2
 -\frac{1}{8m}\rho^B_0 (\nabla
\theta_{\phi} )^2 \nonumber \\
&&\quad\quad-
\frac{1}{2}\Omega^2(\theta_{\tilde{\Delta}}-\theta_{\phi})^2 - 2
|\phi_{0}| \delta |\phi| \ \dot{\theta}_{\phi} \nonumber\\
&& \quad\quad  + (2\mu - \nu)\frac{U_{\rm eff}}{U}(\delta |\phi|)^2
+ \frac{2g}{U}\delta|\tilde{\Delta}|\delta |\phi| -
\frac{1}{U} (\delta|\tilde{\Delta}|)^2 \bigg\}
\nonumber\\
&+& \int d^4q \int d^4p  \bigg\{ {\cal{M}}_{\theta_{\tilde{\Delta}}
\theta_{\tilde{\Delta}}} (q,p) \, {\tilde\Theta}_{\tilde{\Delta}}(q)
{\tilde\Theta}_{\tilde{\Delta}}(-q) \nonumber \\
&& \quad\quad \quad\quad +{\cal{M}}_{\tilde{\Delta} \tilde{\Delta}}
(p, q) \,
\tilde\delta|\tilde{\Delta}|(q) \tilde\delta|\tilde{\Delta}|(-q) \nonumber\\
&& \quad\quad\quad\quad\quad+ {\cal{M}}_{\theta_{\tilde{\Delta}}
\tilde{\Delta}} (p,q) \, \tilde{\Theta}_{\tilde{\Delta}}(q)
\tilde\delta|\tilde{\Delta}| (-q) \bigg\} \,,\label{Squad}
\end{eqnarray}
for ${\cal M}$-functions to be given later.
In (\ref{Squad}) $\delta |\tilde{\Delta}|= |\tilde{\Delta}| -
|\tilde{\Delta_{0}}| $, with Fourier transform $\tilde\delta |\tilde{\Delta}|$ and $\tilde{\Theta}_{\tilde{\Delta}}$ is the Fourier transform of the Galilean invariant $\Theta_{\tilde{\Delta}}
=\dot{\theta}_{\tilde{\Delta}} + (\nabla \theta_{ \tilde{\Delta}
})^2/ 4m $. The varying coupling strength (as $\nu$ varies with external magnetic field) is $U_{eff}= U+ g^2/(2\mu-\nu)$.  The phases of ${\tilde{\Delta}}$ and $\phi$ are coupled with strength $\Omega^2=(2 g/ U)
|\phi_{0}||\tilde{\Delta}_{0}|$.
Defining $E^2_{p}=\varepsilon_{p}^2 + |{\Delta}_{0}|^2$ and
$\varepsilon_{p}={\bf p}^2/2m-\mu$, the fermion number density is $\rho_0 = \rho^F_0 +
\rho^B_0 $, where $
 \rho^F_0 = \int d^3 {\bf p} / (2\pi)^3 \ \left[ 1 -
\varepsilon_{p}/E_{p}
  \right] $
is the explicit fermion density, and $\rho^B_0 = 2|\phi_{0}|^2$ is
due to molecules (two fermions per molecule). For more details see \cite{rivers}.

The separation of the action (\ref{Squad}) into local and seemingly non-local parts is somewhat misleading.  The ${\cal M}$s have $q$-independent terms that lead to further contributions to the local part of the action (the effective local Lagrangian). These are proportional to $\dot{\theta}_{\tilde{\Delta}}^2$ and $(\delta|\tilde{\Delta}|)^2$
and have been displayed elsewhere \cite{rivers}.  The remainders of the double integrals comprise different  non-local combinations of two-Fermion cuts. They are the source of the Landau damping, which
occurs in a collisionless regime via direct dissipationless energy
transfer from the collective mode to single particles. For the relatively simple ${\cal M}$s for the problem in hand of a homogeneous system it is easier not to make this separation into local and non-local parts.

The dynamical equations of the collective modes can be
obtained by taking the variation of the above effective action with
respect to the associated field variables. In what follows, we will
consider broad and narrow resonances in turn.

\section{Broad resonance}

In the case of a broad resonance the one-channel model, which is
obtained by eliminating the molecular bosons from the above
two-channel effective action, suffices to describe its
dynamics~\cite{giorgini,sademelo}. Thus, the corresponding action is
given by setting $\phi=0 $ in the action~(\ref{Lin}) which, in turn,
gives $\tilde{\Delta}=\Delta$.

Consider a spatially homogeneous condensate. If, at time $t=0$ we
turn on a small instantaneous homogeneous change in the external
field this will induce a homogeneous perturbation $\delta_{\Delta}$
in the gap parameter (condensate amplitude). On the other hand, since the gradient of
the phase is the fluid velocity, a spontaneous translation (kick) of
the system will induce a perturbation $\delta_{\theta_{\Delta}}$ in
the phase, which we need to include for completeness. Although we
have not included a trap to constrain the condensate in the
analysis, to keep the results as simple as possible, this
translation of the system could be implemented through a kick
on the trap.

The linearized equations of motion are Fourier analyzed in terms of frequency $\omega$ or, more conveniently,
expressed in terms of the Laplace variable $ s = -i\omega$ as
\begin{eqnarray}
- s^2 \, {\cal{M}}_{\theta_{\Delta} \theta_{\Delta}} (s^2) \,
\theta_{\Delta} + s \, {\cal{M}}_{\theta_{\Delta} \Delta} (s^2) \,
\delta|{\Delta}| &=&
\delta_{\theta_{\Delta}} \, , \nonumber \\
- s \, {\cal{M}}_{\theta_{\Delta} \Delta} (s^2) \, \theta_{\Delta} +
{\cal{M}}_{\Delta \Delta} (s^2)\,  \delta|{\Delta}| &=&
\delta_{\Delta} \,
\end{eqnarray}
(where we have dropped the tildes that normally represent transforms).
After some straightforward but tedious calculation the $ {\cal{M}}s$ are found to be
\begin{eqnarray}
{\cal{M}}_{\theta_{\Delta} \theta_{\Delta}} (s^2) &=& \int \frac{d^3
{\bf{p}}}{ ( 2\pi)^3} \, \frac{1}{E_{p}} \frac{ |\Delta_{0}|^2}{s^2
+ 4 E^2_{p}} \, , \nonumber\\
 {\cal{M}}_{\theta_{\Delta}
\Delta} (s^2) &=& \int \frac{d^3 {\bf{p}}}{ ( 2\pi)^3} \,
\frac{\varepsilon_{p}}{E_{p}} \frac{2 |\Delta_{0}|}{s^2
+ 4 E^2_{p}} \, , \nonumber\\
{\cal{M}}_{\Delta \Delta} (s^2) &=& \int \frac{d^3 {\bf{p}}}{ (
2\pi)^3} \, \left[ \frac{\varepsilon^2_{p}}{E_{p}}  \frac{4 }{s^2 +
4 E^2_{p}} - \frac{1}{ E_{p}} \right].
\end{eqnarray}

 \begin{figure}
\centering
\includegraphics[width=0.7\columnwidth]{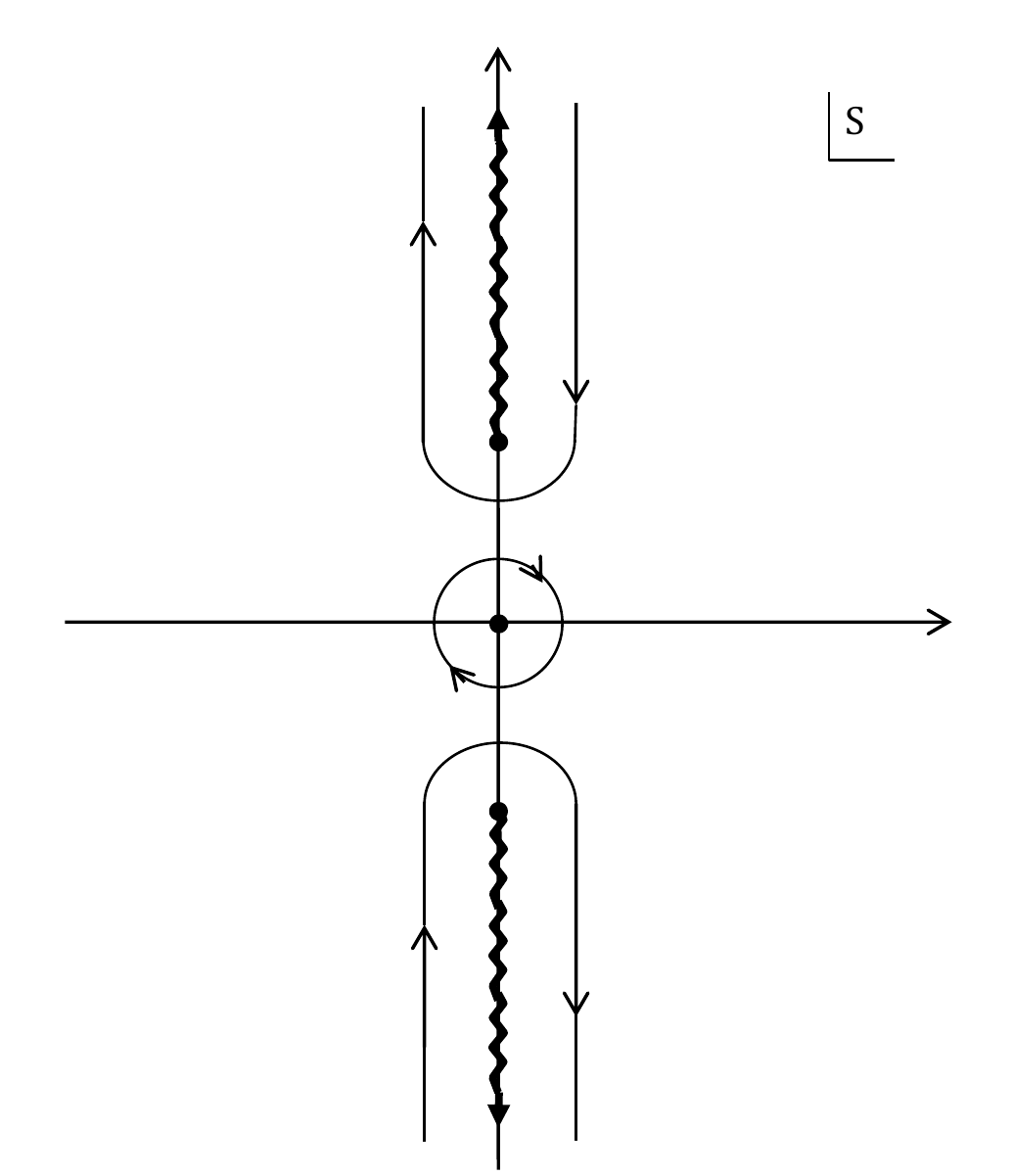}
 \caption{This Fig. shows the analytic structure of the two-Fermion amplitudes discussed in the text in the complex-s plane. The contour is deformed from the Bromwich contour that encompasses
 all singularities in a counterclockwise way. In addition to the pole at $s=i0$ due to the phonon, the cuts are displayed by a wiggly line.} \label{Fig1}
\end{figure}
The solutions of the Laplace transform of the equations are
\begin{eqnarray}
\theta_{\Delta} (s) &=& \frac{ -{\cal{M}}_{\Delta \Delta} (s^2) \,
\delta_{\theta_{\Delta}} + s \, {\cal{M}}_{\theta_{\Delta} \Delta}
(s^2) \, \delta_{\Delta}}{ s^2 \, {\cal{D}} (s^2)} \, , \nonumber
\\
\delta|{\Delta}| (s) &=& \frac{ s^2 \, {\cal{M}}_{\theta_{\Delta}
\theta_\Delta} (s^2) \delta_{{\Delta}} - s \,
{\cal{M}}_{\theta_{\Delta} \Delta} (s^2) \delta_{\theta_{\Delta}}}{
s^2 \, {\cal{D}} (s^2)} \, , \label{theta2}
\end{eqnarray}
where
\begin{equation}
{\cal{D}} (s^2) = {\cal{M}}_{\theta_{\Delta} \theta_\Delta} (s^2)
{\cal{M}}_{\Delta \Delta} (s^2)- {\cal{M}}^2_{\theta_{\Delta}
\Delta} (s^2) \, . \label{D_def}
\end{equation}
The real-time evolution of $\theta (t)$ and $ \delta|{\Delta}| (t) $
can be obtained from carrying out the inverse Laplace transform
along the contour in the s-plane shown in Fig.1~\cite{wang}.

The
singularities include the pole of the phonon mode at $s=i 0$ as well
as the branch cuts, extending from $s=\pm i E_{\rm th} $ to $s=\pm
i \infty$ due to the possibility of breaking a condensate into
fermionic excitations. The solutions to equations (\ref{theta2}) are obtained as
\begin{widetext}
\begin{eqnarray}
\theta_{\Delta} (t) &=& \frac{{\cal{M}}_{\theta_{\Delta} \Delta}
(0)}{ {\cal{D}} (0)}  \delta_{\Delta} - \frac{2}{\pi} \int_{E_{{\rm
{th}}}}^{\infty} \frac{ d \omega} {\omega^2} \frac{ {\cal{D}}_{R}
(\omega) {\cal{M}}_{\Delta \Delta \, I} (\omega) - {\cal{D}}_{I}
(\omega) {\cal{M}}_{\Delta \Delta \, R} (\omega)}{{\cal{D}}_{R}^2
(\omega) + {\cal{D}}_{I}^2 (\omega)} \sin [\omega t]
\delta_{\theta_\Delta}
\nonumber\\
&& \quad\quad\quad\quad \quad\quad\quad\quad+ \frac{2}{\pi}
\int_{E_{{\rm {th}}}}^{\infty} \frac{ d \omega} {\omega} \frac{
{\cal{D}}_{R} (\omega) {\cal{M}}_{\Delta \theta_{\Delta} \, I}
(\omega) - {\cal{D}}_{I} (\omega) {\cal{M}}_{\Delta \theta_{\Delta}
\, R} (\omega)}{{\cal{D}}_{R}^2 (\omega) +{\cal{D}}_{I}^2 (\omega)}
\sin [\omega t] \delta_{{\Delta}} \, ,  \\
 \delta|{\Delta}|
(t) &=& -\frac{{\cal{M}}_{\theta_{\Delta} \Delta} (0)}{ {\cal{D}}
(0)} \delta_{\theta_{\Delta}}  -\frac{2}{\pi} \int_{E_{{\rm
{th}}}}^{\infty}  d \omega  \frac{ {\cal{D}}_{R} (\omega)
{\cal{M}}_{\theta_{\Delta} \theta_{\Delta} \, I} (\omega) -
{\cal{D}}_{I} (\omega) {\cal{M}}_{\theta_{\Delta} \theta_{\Delta} \,
R} (\omega)}{{\cal{D}}_{R}^2 (\omega) +{\cal{D}}_{I}^2 (\omega)}
\sin [\omega t] \delta_{\Delta}
\nonumber\\
&& \quad\quad\quad\quad \quad\quad\quad\quad-  \frac{2}{\pi}
\int_{E_{{\rm {th}}}}^{\infty} \frac{ d \omega} {\omega} \frac{
{\cal{D}}_{R} (\omega) {\cal{M}}_{\theta_{\Delta} \Delta \, I}
(\omega) - {\cal{D}}_{I} (\omega) {\cal{M}}_{\theta_{\Delta} \Delta
\, R} (\omega)}{{\cal{D}}_{R}^2 (\omega) +{\cal{D}}_{I}^2 (\omega)}
\sin [\omega t] \delta_{\theta_\Delta} \, . \label{theta_t}
\end{eqnarray}
\end{widetext}
 The first terms of the above results come from the
 contribution
 of
the pole  and the remaining terms are due to the contributions of  the cuts.
All of the $\cal{M}$s and ${\cal{D}}$ acquire an imaginary part as
$s=i\omega$ crosses the cuts, where their real and imaginary parts
are denoted by
\begin{eqnarray}
{\cal{M}} (s=i \omega \mp \epsilon) &=&  {\cal{M}}_{R} (\omega)
\pm i {\cal{M}}_{I} (\omega) \, ,\nonumber\\
 {\cal{D}} (s=i
\omega \mp \epsilon) &=& {\cal{D}}_{R} (\omega) \pm i {\cal{D}}_{I}
(\omega) \, . \label{cut}
\end{eqnarray}
From Eq.~(\ref{D_def}) ${\cal{D}}_{R,I}$ are found to be
\begin{eqnarray}
 {\cal{D}}_{R} (\omega) &=& \left[ {\cal{M}}_{\theta_{\Delta} \theta_{\Delta} \, R}  {\cal{M}}_{\Delta \Delta \, R}
 -
  {\cal{M}}_{\theta_{\Delta} \theta_{\Delta} \, I}  {\cal{M}}_{\Delta \Delta \, I}  \right.\nonumber\\
  &&\quad\quad\quad \left.-{\cal{M}}^2_{\theta_{\Delta} \Delta \, R}\
  +
   {\cal{M}}^2_{\theta_{\Delta} \Delta \, I}  \right] (\omega) \nonumber \\
  {\cal{D}}_{I} (\omega) &=& \left[ {\cal{M}}_{\theta_{\Delta} \theta_{\Delta} \, R}  {\cal{M}}_{\Delta \Delta \, I}+
  {\cal{M}}_{\theta_{\Delta} \theta_{\Delta} \, I}  {\cal{M}}_{\Delta \Delta \, R}  \right.\nonumber\\
 &&\quad\quad\quad \left. - 2 {\cal{M}}_{\theta_{\Delta} \Delta \, R}  {\cal{M}}_{\theta_{\Delta} \Delta \, I} \right] (\omega)\, .
   \label{D_RI}
\end{eqnarray}
Upon  turning on the perturbation $\delta_{\theta}$
($\delta_{\Delta}$), since the phase mode is generally coupled to
the amplitude mode, it can drive not only the phase (amplitude)
modes but also the amplitude (phase) modes away from their initial
equilibrium values. Because of this coupling the perturbed modes relax to their
respective saturated values given above. However, since the phase fluctuations can couple to the density
fluctuations, we can observe this collective phase mode from the
spectrum of the density-density correlation function (see
Ref.~\cite{ohashi} for details). Here we merely focus on the
real-time behavior of the collective amplitude mode that can be
observed by experimentally time-resolved techniques~\cite{chin}.
The main results below follow directly from Eqs.(\ref{theta_t}).

\subsection{The BCS regime} In the BCS regime for $\mu
>0$, the threshold energy is $E_{\rm th}= 2 |{\Delta_{0}}|$.
The real and imaginary parts of the ${\cal{M}}$s are summarized as
follows:
\begin{eqnarray}
{\cal{M}}_{\theta_{\Delta} \theta_{\Delta} \, R} (\omega)&=&
\frac{m}{ 2\pi^2} (2 m \epsilon_F)^{\frac{1}{2}}
|{\hat{\Delta}_{0}}|^2 I_1 (\hat{\omega}, |\hat{\mu}|,
|{\hat{\Delta}_{0}}|) \, ; \nonumber
 \\
{\cal{M}}_{\theta_{\Delta} \theta_{\Delta} \, I} (\omega) &=&
\frac{m}{ 2\pi^2} (2 m \epsilon_F)^{\frac{1}{2}}
\frac{|{\hat{\Delta}_{0}}|}{4 \,|\hat{\omega}| \, \sqrt{
(\hat{\omega}/2)^2-|{\hat{\Delta}_{0}}|^2}}
\nonumber\\
&&\times ( J_{+ \,1} (\hat{\omega}, \hat{\mu}, |{\hat{\Delta}_{0}}|)- J_{+ \, 1}
(-\hat{\omega}, \hat{\mu}, |{\hat{\Delta}_{0}}|))\, ;\nonumber
\\
 {\cal{M}}_{\theta_{\Delta} \Delta \, R} (\omega) &=& \frac{m}{
2\pi^2} (2 m \epsilon_F)^{\frac{1}{2}} |{\hat{\Delta}_{0}}| I_2
(\hat{\omega}, \hat{\mu}, |{\hat{\Delta}_{0}}|) \, ;\nonumber \\
{\cal{M}}_{\, \theta_{\Delta} \, \Delta \, I} (\omega) &=& \frac{m}{
2\pi^2} (2 m \epsilon_F)^{\frac{1}{2}} \frac{|{\hat{\Delta}_{0}}|}{2
| \hat{\omega}|} \nonumber \\
&& \times( J_{-\, 1}  (\hat{\omega}, \hat{\mu}, |{\hat{\Delta}_{0}}|)- J_{- \, 1} (-\hat{\omega}, \hat{\mu}, |{\hat{\Delta}_{0}}|))\, ;\nonumber \\
{\cal{M}}_{\Delta  \Delta  R (I)}  (\omega) &=& 4 \left(
\frac{ \hat{\omega}^2}{4 |{\hat{\Delta}_{0}}|^2}-1 \right)  \,
{\cal{M}}_{\theta_{\Delta} \theta_{\Delta} \, R \,(I)}  (\omega)\, ;
\nonumber \, ,
\end{eqnarray}
where
\begin{widetext}
\begin{eqnarray}
&& I_1 (\hat{\omega}, \hat{\mu}, |{\hat{\Delta}_{0}}|) ={\cal{P}}
\int_0^{\infty} dx \frac{-
x^2}{\sqrt{(x-{\hat{\mu}})^2+|{\hat{\Delta}_{0}}|^2} \,
\sqrt{\hat{\omega}^2-4[(x-{\hat{\mu}})^2+|{\hat{\Delta}_{0}}|^2]}
} \, ,   \label{I_1} \\
&& J_{\pm \,1} (\hat{\omega}, \hat{\mu}, |{\hat{\Delta}_{0}}|) = - \pi
\theta[\hat{\omega}-2|{\hat{\Delta}_{0}}|] \left\{ \left[ \sqrt{
(\hat{\omega}/2)^2-|{\hat{\Delta}_{0}}|^2 }
+\hat{\mu}\right]^{\frac{1}{2}}
\pm \theta[2\sqrt{\hat{\mu}^2+|{\hat{\Delta}_{0}}|^2}
-\hat{\omega} ] \left[\hat{\mu}- \sqrt{(\hat{\omega}/2)^2-|{\hat{\Delta}_{0}}|^2
} \right]^{\frac{1}{2}} \right\}  \label{J_1} \nonumber \\
\\
&& I_2 (\hat{\omega}, \hat{\mu}, |{\hat{\Delta}_{0}}|)={\cal{P}}
\int_0^{\infty} dx \frac{- 2 x^{\frac{1}{2}} (x-\hat{\mu})
}{\sqrt{(x-{\hat{\mu}})^2+|{\hat{\Delta}_{0}}|^2} \,
\sqrt{\hat{\omega}^2-4[(x-{\hat{\mu}})^2+|{\hat{\Delta}_{0}}|^2]}
}  \label{I_2}
 \end{eqnarray}
 \end{widetext}
 We define
$\hat{\omega}=\omega/\epsilon_F,\hat{\mu}=\mu/\epsilon_F,
|{\hat{\Delta}_{0}}|=|{{\Delta}_{0}}|/\epsilon_F$, and the symbol
${\cal{P}}$ means that the principal value of the integral is taken.
The damping dynamics of the collective motion, particularly at late
times, largely depends on the behavior of the ${\cal{M}}(\omega)$s for
$\omega$  near the threshold energy, i.e. $\omega =E_{\rm th}+0^+ $.
These are found to be
\begin{eqnarray}
&& {\cal{M}}_{\theta_{\Delta} \theta_{\Delta} \, R} (\omega=E_{\rm
th}+0^+) \simeq \bar{{\cal{M}}}_{\theta_{\Delta}
\theta_{\Delta} \, R} \, ;\nonumber\\
&& {\cal{M}}_{\theta_{\Delta} \theta_{\Delta} \, I} (\omega=E_{\rm
th}+0^+) \simeq \bar{{\cal{M}}}_{\theta_{\Delta}
\theta_{\Delta} \, I} \left(\frac{\omega}{E_{\rm th}}-1\right)^{-\frac{1}{2}} \, ;\nonumber\\
&& {\cal{M}}_{\theta_{\Delta} {\Delta} \, R} (\omega= E_{\rm th}+0^+
) \simeq \bar{{\cal{M}}}_{\theta_{\Delta}
{\Delta} \, R} \, ; \nonumber\\
&& {\cal{M}}_{\theta_{\Delta} {\Delta} \, I} (\omega=E_{\rm
th}+0^+)\simeq \bar{{\cal{M}}}_{\theta_{\Delta} {\Delta} \,
I}\left(\frac{\omega}{E_{\rm th}}-1\right)^{\frac{1}{2}}\, ; \nonumber\\
&& {\cal{M}}_{\Delta \Delta \, R } (\omega= E_{\rm th}+0^+)\simeq 8
\bar{{\cal{M}}}_{\theta_{\Delta} \theta_{\Delta} \,
R } \left(\frac{\omega}{E_{\rm th}}-1\right) \, ; \nonumber\\
&& {\cal{M}}_{\Delta \Delta \, I } (\omega= E_{\rm th}+0^+)\simeq 8
\bar{{\cal{M}}}_{\theta_{\Delta} \theta_{\Delta} \,
I} \left(\frac{\omega}{E_{\rm th}}-1\right)^{\frac{1}{2}} \,  \nonumber\\
\label{m_th_bcs}
\end{eqnarray}
with
\begin{eqnarray}
&& \bar{{\cal{M}}}_{\theta_{\Delta} \theta_{\Delta} \, R} =-
\frac{m}{
2\pi^2} (2 m \mu)^{\frac{1}{2}}\frac{1}{4} \left( 1+ \frac{\sqrt{\mu^2+ |{{\Delta}_{0}}|^2}}{\mu} \right) \, ; \nonumber\\
&& \bar{{\cal{M}}}_{\theta_{\Delta} \theta_{\Delta} \, I} =
-\frac{m}{ 2\pi^2} (2 m \mu)^{\frac{1}{2}} \frac{\sqrt{2}\pi}{8} \,
;\nonumber\\
 && \bar{{\cal{M}}}_{\theta_{\Delta} {\Delta} \, R}
= \frac{m}{
2\pi^2} (2 m \mu )^{\frac{1}{2}} \frac{1}{4} \ln \left[\frac{|1+\frac{|{{\Delta}_{0}}|}{\sqrt{\mu^2+ |{{\Delta}_{0}}|^2}}|}{|1-\frac{|{{\Delta}_{0}}|}{\sqrt{\mu^2+ |{{\Delta}_{0}}|^2}}|}   \right]\, ;\nonumber \\
&&\bar{{\cal{M}}}_{\theta_{\Delta} {\Delta} \, I}=- \frac{m}{
2\pi^2} (2 m \mu )^{\frac{1}{2}} \frac{\sqrt{2}\pi}{4}
\frac{|{{\Delta}_{0}}|}{\mu} \, .\nonumber\\ \label{m_bar_bcs}
\end{eqnarray}
As expected, ${\cal{M}}_{\theta_\Delta \Delta \,
R}$ and ${\cal{M}}_{\theta_\Delta \Delta \, R}$ vanish in the deep
BCS regime in which $\mu \gg |{{\Delta}_{0}}|$, resulting from
particle-hole symmetry.  The well-known divergence on the BCS
density of the state at threshold energy $E_{\rm th}= 2
|{\Delta_{0}}|$~\cite{giorgini}  renders ${\cal{M}}_{\theta_{\Delta}
\theta_{\Delta} \, I} (\omega=E_{\rm th}+0^+)$ singular. Together
with the behavior of ${\cal{M}}_{\Delta \Delta \, R (I) } (\omega=
E_{\rm th}+0^+)$ given by Eqs.(\ref{m_th_bcs}), which shows their
vanishing at threshold energy,
  it gives ${\cal{D}}_{R,I}$, defined
  in the expressions (\ref{D_RI}),  as follows:
\begin{eqnarray}
&& {\cal{D}}_{R} (\omega=E_{\rm th}+0^+) \simeq
\bar{{\cal{D}}}_{R} \, ;\nonumber\\
 && {\cal{D}}_{I} (\omega=E_{\rm th}+0^+) \simeq \bar{{\cal{D}}}_{I}\left(\frac{\omega}{E_{\rm
th}}-1\right)^{\frac{1}{2}} \, , \label{D_th}
\end{eqnarray}
where
\begin{eqnarray}
&& \bar{{\cal{D}}}_{R} = -8 \bar{{\cal{M}}}^2_{\theta_{\Delta}
\theta_{\Delta} \,
I}-\bar{{\cal{M}}}^2_{\theta_{\Delta} {\Delta} \, R} \, ;\nonumber\\
&&\bar{{\cal{D}}}_{I}= 16 \bar{{\cal{M}}}_{\theta_{\Delta}
\theta_{\Delta} \, R}\bar{{\cal{M}}}_{\theta_{\Delta}
\theta_{\Delta} \, I} -2 \bar{{\cal{M}}}_{\theta_{\Delta} {\Delta}
\, R } \bar{{\cal{M}}}_{\theta_{\Delta} {\Delta} \, I} \, .
\end{eqnarray}

As seen from Eq.(\ref{theta2}), the
pole at $s=i0$ seems fictitious with respect to the amplitude
perturbation~\cite{gur1}, only becoming a true pole when an
additional phase perturbation is turned on. It is the terms involving the singular behaviour of
${\cal{M}}_{\theta_{\Delta} \theta_{\Delta} \, I}$ at $\omega$ near
the threshold energy that give the dominant contributions to   the
late-time dynamics of the amplitude mode.  Thus, one finds that the integrands in
Eq.(\ref{theta_t}) behave as
\begin{eqnarray}
&& \frac{ {\cal{D}}_{R} {\cal{M}}_{\theta_{\Delta} \theta_{\Delta}
\, I}  - {\cal{D}}_{I}  {\cal{M}}_{\theta_{\Delta} \theta_{\Delta}
\, R} }{{\cal{D}}_{R}^2  +{\cal{D}}_{I}^2 }|_{\omega=E_{\rm
th}+0^+} \nonumber\\
&&\quad \quad\quad \quad\quad \simeq
\frac{\bar{{\cal{M}}}_{\theta_{\Delta} \theta_{\Delta} \,
I}}{\bar{{\cal{D}}}_{R}} \left(\frac{\omega}{E_{\rm
th}}-1\right)^{-\frac{1}{2}}\, \label{I1_th_bcs_broad}\\
 && \frac{ {\cal{D}}_{R}
{\cal{M}}_{\theta_{\Delta} \Delta \, I} - {\cal{D}}_{I}
{\cal{M}}_{\theta_{\Delta} \Delta \, R} }{{\cal{D}}_{R}^2
+{\cal{D}}_{I}^2}|_{\omega=E_{\rm
th}+0^+} \nonumber\\
&&\quad \quad\quad \quad\simeq
\frac{\bar{{\cal{D}}}_{R}\bar{{\cal{M}}}_{\theta_{\Delta} {\Delta}
\, I}-\bar{{\cal{D}}}_{I}\bar{{\cal{M}}}_{\theta_{\Delta} {\Delta}
\, R}}{\bar{{\cal{D}}}^2_{R}} \left(\frac{\omega}{E_{\rm
th}}-1\right)^{\frac{1}{2}}\, \label{I2_th_bcs_broad}.
\end{eqnarray}
These two terms in turn will lead to the different damping behaviour
after carrying out the integration over $\omega$.

Putting all this together, we find the late-time solutions as
\begin{widetext}
\begin{eqnarray}
 \delta|{\Delta}|
(t) &\simeq& -\frac{{\cal{M}}_{\theta_{\Delta} \Delta} (0)}{
{\cal{D}} (0)} \delta_{\theta_{\Delta}}  -\frac{2}{\sqrt{\pi}}
\frac{\bar{{\cal{M}}}_{\theta_{\Delta} \theta_{\Delta} \,
I}}{\bar{{\cal{D}}}_{R}} \frac{E_{\rm th} }{E_{\rm th}^{1/2} \,
t^{1/2}} \cos[ E_{\rm th} t-\pi/4] \delta_{\Delta}
\nonumber\\
&&\quad\quad\quad \quad\quad\quad\quad\quad-\frac{1}{\sqrt{\pi}}
\frac{\bar{{\cal{D}}}_{R}\bar{{\cal{M}}}_{\theta_{\Delta} {\Delta}
\, I}-\bar{{\cal{D}}}_{I}\bar{{\cal{M}}}_{\theta_{\Delta} {\Delta}
\, R}}{\bar{{\cal{D}}}^2_{R}} \frac{1}{E_{\rm th}^{3/2}\, t^{3/2}}
\cos[ E_{\rm th} t+ \pi/4] \delta_{\theta_\Delta} \, .
\label{delta_damp}
\end{eqnarray}
\end{widetext}
 In the BCS regime, $ \delta|{\Delta}| (t) $ decays at late times dominantly as $t^{-1/2}$~\cite{vol,gur1}
 due to the amplitude perturbation  while it has a subdominant decay as
$t^{-3/2}$ given by the perturbation $\delta_{\theta}$.

Additionally, if it were possible to implement small phase perturbations, $ \delta|{\Delta}| (t) $
 decays to a non-zero value proportional to the perturbation
 $\delta_{\theta}$, resulting from the amplitude-phase coupling. This looks to give a shift on the saturated value
 that the amplitude mode decays to from its initial value, but this becomes vanishingly small in the deep BCS regime where ${\cal{M}}_{\theta_{\Delta} \Delta}\simeq 0 $. We might hope that this subdominant behaviour could be made visible by judicious choice of initial perturbations, but we do not know.

\subsection{The BEC regime}

In the BEC regime the chemical
potential changes its sign, $\mu <0$. The threshold energy now
becomes $E_{\rm th} = \sqrt{ 2 ( |\mu |^2 + | \Delta_0 |^2 )}$. The
real and imaginary part of the ${\cal M}$s obtained from Eqs.(\ref{cut}) are
given by
\begin{eqnarray}
&& {\cal{M}}_{\theta_{\Delta} \theta_{\Delta} \, R} (\omega)=
\frac{m}{ 2\pi^2} (2 m \epsilon_F)^{\frac{1}{2}}
|{\hat{\Delta}_{0}}|^2 I_1 (\hat{\omega},-| \hat{\mu}|,
|{\hat{\Delta}_{0}}|) \, ; \nonumber
 \\
&& {\cal{M}}_{\theta_{\Delta} \theta_{\Delta} \, I} (\omega) =
\frac{m}{ 2\pi^2} (2 m \epsilon_F)^{\frac{1}{2}}
\frac{|{\hat{\Delta}_{0}}|}{4 |\hat{\omega}| \, \sqrt{
(\hat{\omega}/2)^2-|{\hat{\Delta}_{0}}|^2}}
\nonumber\\
&&\quad\quad \quad\times ( J_2 (\hat{\omega}, -|\hat{\mu}|,
|{\hat{\Delta}_{0}}|)- J_2 (-\hat{\omega}, -|\hat{\mu}|,
|{\hat{\Delta}_{0}}|))\, ;\nonumber
\\
 &&{\cal{M}}_{\theta_{\Delta} \Delta \, R} (\omega) = \frac{m}{
2\pi^2} (2 m \epsilon_F)^{\frac{1}{2}} |{\hat{\Delta}_{0}}| I_2
(\hat{\omega}, - |\hat{\mu}|, |{\hat{\Delta}_{0}}|) \, ;\nonumber \\
&&{\cal{M}}_{\, \theta_{\Delta} \, \Delta \, I} (\omega) = \frac{m}{
2\pi^2} (2 m \epsilon_F)^{\frac{1}{2}} \frac{|{\hat{\Delta}_{0}}|}{2
|\hat{\omega}|} \nonumber \\
&&\quad \quad\quad\times( J_2 (\hat{\omega}, -|\hat{\mu}|, |{\hat{\Delta}_{0}}|)- J_2 (-\hat{\omega},-| \hat{\mu}|, |{\hat{\Delta}_{0}}|))\, ;\nonumber \\
&& {\cal{M}}_{\, \Delta \,\, \Delta \, R \, (I)} (\omega) = 4 \left(
\frac{ \hat{\omega}^2}{4 |{\hat{\Delta}_{0}}|^2}-1 \right) \,
{\cal{M}}_{\theta_{\Delta} \theta_{\Delta} \, R\, (I)} (\omega)\,
,\nonumber\\
\end{eqnarray}
where
\begin{eqnarray}
J_2 (\hat{\omega}, \hat{\mu}, |{\hat{\Delta}_{0}}|) &=& - \pi
\theta[\hat{\omega}-2
\sqrt{ |\mu |^2 + | \Delta_0 |^2}  ]  \nonumber\\
&&\quad \times \left[\sqrt{ (\hat{\omega}/2)^2-|{\hat{\Delta}_{0}}|^2
} -|\hat{\mu}|\right]^{\frac{1}{2}}
\end{eqnarray}
together with Eqs.(\ref{I_1}),(\ref{I_2}).  For $\omega$ near the
threshold energy, the ${\cal{M}}$s can be simplified as
\begin{eqnarray}
&& {\cal{M}}_{\theta_{\Delta} \theta_{\Delta} \, R} (\omega=E_{\rm
th}+0^+) \simeq \bar{{\cal{M}}}_{\theta_{\Delta}
\theta_{\Delta} \, R} \, ;\nonumber\\
&& {\cal{M}}_{\theta_{\Delta} \theta_{\Delta} \, I} (\omega=E_{\rm
th}+\epsilon) \simeq \bar{{\cal{M}}}_{\theta_{\Delta}
\theta_{\Delta} \, I} \left(\frac{\omega}{E_{\rm th}}-1\right)^{\frac{1}{2}} \, ; \nonumber\\
&& {\cal{M}}_{\theta_{\Delta} {\Delta} \, R} (\omega= E_{\rm th}+0^+
) \simeq \bar{{\cal{M}}}_{\theta_{\Delta}
{\Delta} \, R} \, ; \nonumber\\
&& {\cal{M}}_{\theta_{\Delta} {\Delta} \, I} (\omega=E_{\rm
th}+0^+)\simeq \bar{{\cal{M}}}_{\theta_{\Delta} {\Delta} \,
I}\left(\frac{\omega}{E_{\rm th}}-1\right)^{\frac{1}{2}}\, ; \nonumber\\
&& {\cal{M}}_{\Delta \Delta \, R} (\omega=E_{\rm th}+\epsilon)\simeq
4  \frac{| \mu |^2}{|{{\Delta}_{0}}|^2} \bar{{\cal{M}}}_{\theta_{\Delta} \theta_{\Delta} \, R}  \, ;
\nonumber\\
&& {\cal{M}}_{\Delta \Delta \, I} (\omega= E_{\rm th}+0^+) \simeq 4  \frac{| \mu |^2}{|{{\Delta}_{0}}|^2}
\bar{{\cal{M}}}_{\theta_{\Delta} \theta_{\Delta} \, I}
\left(\frac{\omega}{E_{\rm th}}-1\right)^{\frac{1}{2}} \, \nonumber \\
\label{m_th_bec_broad}
\end{eqnarray}
which, in the deep BEC regime, becomes
\begin{eqnarray}
&& \bar{{\cal{M}}}_{\theta_{\Delta} \theta_{\Delta} \, R}  \simeq   \frac{m}{ 2\pi^2} (2 m \mu)^{\frac{1}{2}}
\frac{(2-\sqrt{2}) \pi}{8}
\frac{|{{\Delta}_{0}}|^2}{|\mu|^2}  \, ;\nonumber\\
 && \bar{{\cal{M}}}_{\theta_{\Delta} \theta_{\Delta} \, I} \simeq
-\frac{m}{ 2\pi^2} (2 m \mu)^{\frac{1}{2}} \frac{\pi}{8}
\frac{|{{\Delta}_{0}}|^2}{|\mu|^2}  \,
;\nonumber\\
 && \bar{{\cal{M}}}_{\theta_{\Delta} {\Delta} \, R}
\simeq \frac{m}{ 2\pi^2} (2 m |\mu| )^{\frac{1}{2}}
\frac{\pi}{2 \sqrt{2}} \frac{|{{\Delta}_{0}}|}{|\mu|} \, ; \nonumber\\
&&\bar{{\cal{M}}}_{\theta_{\Delta} {\Delta} \, I}\simeq - \frac{m}{
2\pi^2} (2 m \mu )^{\frac{1}{2}} \frac{\pi}{4}
\frac{|{{\Delta}_{0}}|}{|\mu|} \, . \label{m_bar_bec_broad}
\end{eqnarray}
Although the behaviour of the above $\cal{M}$s seems different from
that in the BCS regime,  $\cal{D}_{R (I)}$ follows Eq.(\ref{D_th})
with $\bar{\cal{D}}_{R (I)}$ given respectively by
\begin{eqnarray}
&& \bar{{\cal{D}}}_{R} = 4  \frac{| \mu |^2}{|{{\Delta}_{0}}|^2} \bar{{\cal{M}}}^2_{\theta_{\Delta}
\theta_{\Delta} \,
I}-\bar{{\cal{M}}}^2_{\theta_{\Delta} {\Delta} \, R} \, ;\nonumber\\
&&\bar{{\cal{D}}}_{I}= 8 \frac{| \mu |^2}{|{{\Delta}_{0}}|^2}
\bar{{\cal{M}}}_{\theta_{\Delta} \theta_{\Delta} \,
R}\bar{{\cal{M}}}_{\theta_{\Delta} \theta_{\Delta} \, I} -2
\bar{{\cal{M}}}_{\theta_{\Delta} {\Delta} \, R }
\bar{{\cal{M}}}_{\theta_{\Delta} {\Delta} \, I} \, .
\label{D_bar_bec_broad}
\end{eqnarray}
So, as $\omega$ is near the threshold energy, the integrands in Eq.(\ref{theta_t}) can be approximated by
\begin{eqnarray}
&& \frac{ {\cal{D}}_{R} {\cal{M}}_{\theta_{\Delta} \theta_{\Delta}
\, I}  - {\cal{D}}_{I}  {\cal{M}}_{\theta_{\Delta} \theta_{\Delta}
\, R} }{{\cal{D}}_{R}^2  +{\cal{D}}_{I}^2 }|_{\omega=E_{\rm
th}+0^+} \nonumber\\
&&\quad \quad\quad  \simeq
\frac{\bar{{\cal{D}}}_{R}\bar{{\cal{M}}}_{\theta_{\Delta}
\theta_{\Delta} \,
I}-\bar{{\cal{D}}}_{I}\bar{{\cal{M}}}_{\theta_{\Delta}
\theta_{\Delta} \, R}}{\bar{{\cal{D}}}^2_{R}}
\left(\frac{\omega}{E_{\rm
th}}-1\right)^{\frac{1}{2}}\, \\
 && \frac{ {\cal{D}}_{R}
{\cal{M}}_{\theta_{\Delta} \Delta \, I}  - {\cal{D}}_{I}
{\cal{M}}_{\theta_{\Delta} \Delta \, R} }{{\cal{D}}_{R}^2
+{\cal{D}}_{I}^2}|_{\omega=E_{\rm
th}+0^+} \nonumber\\
&&\quad \quad \simeq
\frac{\bar{{\cal{D}}}_{R}\bar{{\cal{M}}}_{\theta_{\Delta} {\Delta}
\, I}-\bar{{\cal{D}}}_{I}\bar{{\cal{M}}}_{\theta_{\Delta} {\Delta}
\, R}}{\bar{{\cal{D}}}^2_{R}}
 \left(\frac{\omega}{E_{\rm
th}}-1\right)^{\frac{1}{2}}\, .
\end{eqnarray}
The same damping behaviour will be seen from the above two respective
contributions.

 Putting all these results together now gives the late-time result:
\begin{widetext}
\begin{eqnarray}
\delta|{\Delta}| (t) &\simeq& -\frac{{\cal{M}}_{\theta_{\Delta}
\Delta} (0)}{ {\cal{D}} (0)} \delta_{\theta_{\Delta}}
-\frac{1}{\sqrt{\pi}}
\frac{\bar{{\cal{D}}}_{R}\bar{{\cal{M}}}_{\theta_{\Delta}
\theta_{\Delta} \,
I}-\bar{{\cal{D}}}_{I}\bar{{\cal{M}}}_{\theta_{\Delta}
\theta_{\Delta} \, R}}{\bar{{\cal{D}}}^2_{R}} \frac{1 }{E_{\rm
th}^{3/2}\, t^{3/2}} \cos[ E_{\rm th} t+\pi/4] \delta_{\Delta}
\nonumber\\
&&\quad\quad\quad \quad\quad\quad\quad\quad-\frac{1}{\sqrt{\pi}}
\frac{\bar{{\cal{D}}}_{R}\bar{{\cal{M}}}_{\theta_{\Delta} {\Delta}
\, I}-\bar{{\cal{D}}}_{I}\bar{{\cal{M}}}_{\theta_{\Delta} {\Delta}
\, R}}{\bar{{\cal{D}}}^2_{R}} \frac{1}{E_{\rm th}^{3/2}\, t^{3/2}}
\cos[ E_{\rm th} t+ \pi/4] \delta_{\theta_\Delta} \, .
\label{decay_bec_broad}
\end{eqnarray}
\end{widetext}
Note that $E_{\rm th}=\sqrt{ 2 ( |\mu |^2 + | \Delta_0 |^2)} $ in
the BEC regime. The amplitude mode decays as $t^{-3/2}$ in both
oscillating terms, but we note that the coefficient of
$\delta_{\theta}$ is the larger in the deep BEC regime, as seen from
Eqs.(\ref{m_bar_bec_broad}) in which
$\bar{{\cal{M}}}_{\theta_{\Delta} {\Delta}} \gg
\bar{{\cal{M}}}_{\theta_{\Delta} {\theta_\Delta}}$ for $ |\mu| \gg
|{{\Delta}_{0}}|$. Insofar as it is possible to implement an initial
phase shift $\delta|{\Delta}| (t)$ then decays to a saturated value
with a nonzero shift from the initial condition. The significance of
introducing the phase perturbation is seen from the two results
above.

In summary, our calculation of the generic $t^{-1/2}$ damping in the BCS superconductor regime and the $t^{-3/2}$ damping in the BEC molecular regime reproduces the results of ~\cite{gur1} obtained by independent methods. As in ~\cite{gur1}, we have seen that the transition between the two regimes of the different
damping behaviours occurs at $\mu=0$. However, for $\delta_{\theta_{\Delta}}\neq 0$, there is a residual displacement of $\delta|{\Delta}| (t)$ in the BEC regime.

\section{Narrow resonance}

The situation is rather different when the Feshbach resonance of
bound bosons is very narrow. We now have a two-channel model in which the
resonance  has to be taken into account explicitly.  To illustrate
this, for simplicity we consider the idealised case~\cite{andreev} of $U=0$ in the above
Lagrangian where the effects from the background scattering length
$a_{\rm bg}$ are ignored as long as the system is away from the deep
BCS regime~\cite{gurarie}.  It leads to $\tilde{\Delta}=- g\phi$
from Eq.(\ref{tildeDelta}), so the field variables of the
collective modes then are the phase $\theta_{\phi}$ and the
amplitude $\delta|\phi|$.

On perturbing the system the collective phenomena (in the same notation as before) are described by the
following equations:

\begin{eqnarray}
- s^2 \, {\cal{M}}_{\theta_{\phi} \theta_{\phi}} (s^2) \,
\theta_{\phi} + s \, {\cal{M}}_{\theta_{\phi} \phi} (s^2) \,
g \delta|\phi| &=&
\delta_{\theta_{\phi}} \, , \nonumber \\
- s \, {\cal{M}}_{\theta_{\phi} \phi} (s^2) \, \theta_{\phi} +
{\cal{M}}_{\phi \phi} (s^2)\, g \delta|\phi| &=& \delta_{\phi} \, .
\end{eqnarray}
All ${\cal{M}}s $ here can be obtained from those of the broad
resonance together with the identification of $ |\tilde{\Delta}_0|=
g |\phi_0|$ as follows:
\begin{eqnarray}
 {\cal{M}}_{\theta_{\phi} \theta_{\phi}} &=& {\cal{M}}_{\theta_{\Delta}
 \theta_{\Delta}}\, ;\,\, {\cal{M}}_{\theta_{\phi} \phi}=  {\cal{M}}_{\theta_{\Delta}
 \Delta} +2 |\phi_0| \, ; \nonumber\\
 {\cal{M}}_{{\phi} \phi} &=& 2( 2 \mu-\bar{\nu}) +  {\cal{M}}_{\Delta
 \Delta}\, . \label{m_narrow}
\end{eqnarray}
The behaviour of $\cal{M}$s at $\omega$ near the threshold energy can
then be read off from the results in the broad resonance case above
and the corresponding behaviour of the added terms.

The presence of the constant inhomogeneous terms in ${\cal{M}}_{\theta_{\phi} \phi}$ and ${\cal{M}}_{{\phi} \phi}$ in (\ref{m_narrow}) allows for different damping behaviour from that seen previously and we shall see that these differences are realized.

\subsection{The BCS regime} For $\mu
> 0$ in the BCS regime, together with Eqs.(\ref{m_th_bcs}),
${\cal{M}}_{{\phi} \phi \, R}$ and $ {\cal{M}}_{\theta_{\phi} \phi \, R}
$ follow
\begin{eqnarray}
&& {\cal{M}}_{{\phi} \phi \, R} (\omega=E_{\rm th} +0^+) \simeq
\bar{{\cal{M}}}_{{\phi} \phi \, R} =2( 2
\mu-\bar{\nu}) \, ; \label{m_bar_phi}\\
&& {\cal{M}}_{\theta_{\phi} \phi \, R} (\omega=E_{\rm th}
+0^+)\simeq\bar{{\cal{M}}}_{\theta_{\phi} \phi \, R}
\nonumber\\
&& \quad =  \frac{m}{
2\pi^2} (2 m \mu )^{\frac{1}{2}} \frac{1}{4} \ln \left[\frac{|1+\frac{g |{{\phi}_{0}}|}{\sqrt{\mu^2+ g^2 |{{\phi}_{0}}|^2}}|}{|1-\frac{g |{{\phi}_{0}}|}{\sqrt{\mu^2+ g^2 |{{\phi}_{0}}|^2}}|}   \right]\,+2 |\phi_0| \,
,\nonumber\\
 \label{m_bar_narrow}
\end{eqnarray}
whereas the behaviour of the other $\cal{M}$s remains the same as in
Eqs.(\ref{m_th_bcs}) and (\ref{m_bar_bcs}) in which the labels
$\theta_{\Delta}$ and $\Delta$ are replaced respectively by the
corresponding $\theta_{\theta}$ and $\phi$, and $ |{\Delta}_0|$ is
changed to $g |\phi_0|$.  For $ 2 \mu \neq
\bar{\nu}$ we have
${\cal{M}}_{{\phi} \phi \, R} (\omega=E_{\rm th} +0^+)\neq 0$. This is crucial and leads to the
rather different results for ${\cal{D}}_{R,I}$, which are obtained
from Eqs.(\ref{D_RI}) in terms of  $\theta_{\phi}$ and $\delta\phi$,
as
\begin{eqnarray}
&& {\cal{D}}_{R} (\omega=E_{\rm th}+0^+) \simeq
\bar{{\cal{D}}}_{R} \, ;\nonumber\\
 && {\cal{D}}_{I} (\omega=E_{\rm th}+0^+) \simeq \bar{{\cal{D}}}_{I}\left(\frac{\omega}{E_{\rm
th}}-1\right)^{-\frac{1}{2}} \,  \label{D_th_narrow}
\end{eqnarray}
with
\begin{eqnarray}
&& \bar{{\cal{D}}}_{R} = \bar{{\cal{M}}}_{\theta_\phi \theta_\phi \,
R} \bar{{\cal{M}}}_{\phi \phi \, R} -8
\bar{{\cal{M}}}^2_{\theta_\phi \theta_\phi \,
I}-\bar{{\cal{M}}}^2_{\theta_{\phi} {\phi} \, R} \, ;\nonumber\\
&&\bar{{\cal{D}}}_{I}=  \bar{{\cal{M}}}_{\theta_{\phi} \theta_{\phi}
\, I}\bar{{\cal{M}}}_{\phi \phi \, R}  \, .
\end{eqnarray}

In particular, in addition to
${\cal{M}}_{\theta_{\phi} \theta_{\phi}\, I }$, ${\cal{D}}_{I}$ now
becomes singular as $\omega$ is near the threshold energy. Thus, the
terms involving either ${\cal{M}}_{\theta_{\phi} \theta_{\phi}\, I
}$ or ${\cal{D}}_{I}$ will become dominant in determining the
late-time dynamics of the amplitude mode. We then find that the integrands in the
corresponding equation to Eq.(\ref{theta_t}), in terms of the
variables $\theta_\phi$ and $\delta \phi$, can be given by
\begin{eqnarray}
&& \frac{ {\cal{D}}_{R} {\cal{M}}_{\theta_{\phi} \theta_{\phi} \, I}
- {\cal{D}}_{I}  {\cal{M}}_{\theta_{\phi} \theta_{\phi} \, R}
}{{\cal{D}}_{R}^2  +{\cal{D}}_{I}^2 }|_{\omega=E_{\rm
th}+0^+} \nonumber\\
&&\quad \quad\quad  \simeq
\frac{\bar{{\cal{D}}}_{R}\bar{{\cal{M}}}_{\theta_{\phi}
\theta_{\phi} \,
I}-\bar{{\cal{D}}}_{I}\bar{{\cal{M}}}_{\theta_{\phi} \theta_{\phi}
\, R}}{\bar{{\cal{D}}}^2_{I}} \left(\frac{\omega}{E_{\rm
th}}-1\right)^{\frac{1}{2}}\, \\
 && \frac{ {\cal{D}}_{R}
{\cal{M}}_{\theta_{\phi} \phi \, I}  - {\cal{D}}_{I}
{\cal{M}}_{\theta_{\phi} \phi\, R}}{{\cal{D}}_{R}^2
+{\cal{D}}_{I}^2}|_{\omega=E_{\rm
th}+0^+} \nonumber\\
&&\quad \quad \quad \quad\quad \quad\simeq
-\frac{\bar{{\cal{M}}}_{\theta_{\phi} {\phi} \,
R}}{\bar{{\cal{D}}}_{I}}
 \left(\frac{\omega}{E_{\rm
th}}-1\right)^{\frac{1}{2}}\, .
\end{eqnarray}

In general, the damping behaviour of the amplitude mode at late times
 then becomes
\begin{widetext}
\begin{eqnarray}
 g \delta|{\phi}|
(t) &\simeq& -\frac{{\cal{M}}_{\theta_{\phi} \phi} (0)}{ {\cal{D}}
(0)} \delta_{\theta_{\Delta}}  -\frac{1}{\sqrt{\pi}}
\frac{\bar{{\cal{D}}}_{R}\bar{{\cal{M}}}_{\theta_{\phi}
\theta_{\phi} \,
I}-\bar{{\cal{D}}}_{I}\bar{{\cal{M}}}_{\theta_{\phi} \theta_{\phi}
\, R}}{\bar{{\cal{D}}}^2_{I}} \frac{1}{E_{\rm th}^{3/2}t^{3/2}}
\cos[ E_{\rm th} t-\pi/4] \delta_{\phi}
\nonumber\\
&&\quad\quad\quad \quad\quad\quad\quad\quad+\frac{1}{\sqrt{\pi}}
\frac{\bar{{\cal{M}}}_{\theta_{\phi} {\phi} \,
R}}{\bar{{\cal{D}}}_{I}} \frac{1}{E_{\rm th}^{3/2}\, t^{3/2}} \cos[
E_{\rm th} t+ \pi/4] \delta_{\theta_\phi} \, . \label{delta_damp}
\end{eqnarray}
\end{widetext}
That is, the decay of the amplitude mode in the narrow resonance
follows the power-law  $t^{-3/2}$, prior to reaching a saturated
value in the BCS regime. This is different from the broad resonance
result quoted earlier where the amplitude mode decays dominantly as
$t^{-1/2}$ at late times instead.  The only exception to this decay
behaviour of the amplitude mode arises when the
unitary limit (infinite scattering length) $ 2 \mu =\bar{\nu}$
occurs in the regime of positive chemical
potential~\cite{rivers,rivers1}. We then have ${\cal{M}}_{\phi \phi
\, R } (\omega)$ of (\ref{m_narrow}) vanishing linearly at $\omega$
near the threshold energy $E_{\rm th}$, i.e. ${\cal{M}}_{\phi \phi
\, R } (\omega= E_{\rm th}+0^+)\approx \left(\frac{\omega}{E_{\rm
th}}-1\right) $, which gives the same $t^{-1/2}$ behaviour as its
general counterpart in the broad resonance given in
Eqs.~({\ref{m_th_bcs}).

This finding of $t^{-3/2}$ damping behaviour in the BCS regime seems
to contradict the conclusion in~\cite{gur1} in which the same
damping dynamics is found  in both broad (one-channel model) and
narrow (two-channel model) resonances.  However, our
idealization of narrow resonances for which we have taken the
contact interaction strength $U=0$ in (\ref{Lin}) cannot provide a
valid description in the regime of deep BCS. For a more general
two-channel model with a nonzero $U$,  characterized by the
background scattering length, $U_{eff} \simeq U$ in the deep BSC
regime~\cite{rivers,rivers1}. Thus, the damping behaviour of~\cite{gur1}
will be recovered in the deep
BCS regime
with a power law  $t^{-1/2} $ decay, the same
result as for the one-channel model, although the details are much messier. The outcome would be that, although
this term might have a small coefficient, it has the dominant
long-time behaviour. It should be possible to see the otherwise subdominant $t^{-3/2}$ behaviour in the BSC regime if $ U \ll g^2/(2\mu-\nu)$.


\subsection{The BEC regime}
In the case of negative chemical potential in the BEC regime,  as a
result of Eqs.({\ref{m_narrow}) and the identification of $
|\tilde{\Delta}_0|= g |\phi_0|$, one finds that the corresponding
$\cal{M}$s at $\omega$ near the threshold energy bear similarity to
those in the broad resonance  in Eqs.(\ref{m_th_bec_broad}),
provided their coefficients are changed. The changes are summarized as
follows:
\begin{eqnarray}
&& {\cal{M}}_{{\phi} \phi \, R} (\omega=E_{\rm th} +0^+) \simeq
\bar{{\cal{M}}}_{{\phi} \phi \, R} \nonumber \\
 && \quad \quad=2( 2 \mu-\bar{\nu}) +   \frac{m}{ 2\pi^2} (2 m \mu)^{\frac{1}{2}}
\frac{(2-\sqrt{2}) \pi}{2} \nonumber \\ \\
 && {\cal{M}}_{\theta_{\phi} \phi \, R} (\omega=E_{\rm th}
+0^+)\simeq\bar{{\cal{M}}}_{\theta_{\phi} \phi \, R}
\nonumber\\
&&\quad\quad\quad\quad= \frac{m}{ 2\pi^2} (2 m |\mu|
)^{\frac{1}{2}} \frac{\pi}{2 \sqrt{2}}
\frac{g|{{\phi}_{0}}|}{|\mu|}\, . \label{m_bar_bec_narrow}
\end{eqnarray}
All other ${\cal{M}}$s follow Eqs.(\ref{m_th_bec_broad}) and
(\ref{m_bar_bec_broad}) by changing again the labels
$\theta_{\Delta}$ and $\Delta$ to $\theta_\phi$ and $\phi$.
Additionally,  $ |{\Delta}_0|$ is replaced by $g |\phi_0|$.
Thus, $\cal{D}_{R (I)}$ follows Eq.(\ref{D_th})
with $\bar{\cal{D}}_{R (I)}$, modified from Eqs.(\ref{D_bar_bec_broad}) and is thus given by
\begin{eqnarray}
&& \bar{{\cal{D}}}_{R} = 4  \frac{| \mu |^2}{g^2 |{{\phi}_{0}}|^2} \bar{{\cal{M}}}^2_{\theta_{\phi}
\theta_{\phi} \,
I}-\bar{{\cal{M}}}^2_{\theta_{\phi} {\phi} \, R} \, ;\nonumber\\
&&\bar{{\cal{D}}}_{I}= 8 \frac{| \mu |^2}{g^2 |{{\phi}_{0}}|^2}
\bar{{\cal{M}}}_{\theta_{\phi} \theta_{\phi} \,
R}\bar{{\cal{M}}}_{\theta_{\phi} \theta_{\phi} \, I} -2
\bar{{\cal{M}}}_{\theta_{\phi} {\phi} \, R }
\bar{{\cal{M}}}_{\theta_{\phi} {\phi} \, I} \, .
\label{D_bar_bec_narrow}
\end{eqnarray}

Then the decay of the amplitude mode at late times is given
by Eq.(\ref{decay_bec_broad}), in which all $\bar{\cal{M}}$s and
$\bar{\cal{D}}$s  are
replaced by those in terms of the variables $\theta_{\phi}$ and
$\delta|\phi|$, as follows:
\begin{widetext}
\begin{eqnarray}
g \delta|{\phi}| (t)  &\simeq& -\frac{{\cal{M}}_{\theta_{\phi} \phi}
(0)}{ {\cal{D}} (0)} \delta_{\theta_{\phi}} -\frac{1}{\sqrt{\pi}}
\frac{\bar{{\cal{D}}}_{R}\bar{{\cal{M}}}_{\theta_{\phi}
\theta_{\phi} \,
I}-\bar{{\cal{D}}}_{I}\bar{{\cal{M}}}_{\theta_{\phi} \theta_{\phi}
\, R}}{\bar{{\cal{D}}}^2_{R}} \frac{1 }{E_{\rm th}^{3/2}\, t^{3/2}}
\cos[ E_{\rm th} t+\pi/4] \delta_{\phi}
\nonumber\\
&&\quad\quad\quad \quad\quad\quad\quad\quad-\frac{1}{\sqrt{\pi}}
\frac{\bar{{\cal{D}}}_{R}\bar{{\cal{M}}}_{\theta_{\phi} {\phi} \,
I}-\bar{{\cal{D}}}_{I}\bar{{\cal{M}}}_{\theta_{\phi} {\phi} \,
R}}{\bar{{\cal{D}}}^2_{R}} \frac{1}{E_{\rm th}^{3/2}\, t^{3/2}}
\cos[ E_{\rm th} t+ \pi/4] \delta_{\theta_\phi} \, .
\label{decay_bec_narrow}
\end{eqnarray}
\end{widetext}
The outcome, that the damping behaviour is $t^{-3/2}$, as for broad
resonances, is consistent with the argument in~\cite{gur1} but, yet again, there is the possibility of the condensate relaxing to a displaced value.

\section{Conclusions}
In this paper we determined the time evolution of a condensate of
cold Fermi atoms, whose interaction can be tuned by a Feshbach
resonance throughout the BCS and BEC regimes, in response to (small) non-equilibrium
perturbations of the phase and amplitude modes. Interaction between the
collective modes and the constituent particles is key for our
understanding of
 the relational dynamics, which occurs in a collisionless regime via the Landau damping effect.

In particular, we have focussed on the evolution of the collective
amplitude mode. For a broad resonance the amplitude mode decays as
$t^{-1/2}$ at late times in the BCS regime with positive chemical
potential whereas it decays as $t^{-3/2}$ in the BEC regime with
negative chemical potential. Our conclusions, although obtained by
different means, agree with those in the literature ~\cite{gur1}.
However, when the resonance is very narrow we disagree with
~\cite{gur1}, which sees no difference between the behavior for
broad or narrow resonances. We find that for an idealized narrow
Feshbach resonance the collective mode decays as $t^{-3/2}$
throughout both the BCS and BEC regimes. Nonetheless,
we expect that the deviation from idealized narrow resonances
restores the original longtime behaviour in the deep BCS regime. The
$t^{-3/2}$ behavior can possibly be seen experimentally in the BCS
regime when the effect of the background scattering length can be
largely ignored.
More robustly, beyond the results of ~\cite{gur1}
the condensate amplitude can decay to a value shifted
from its initial condition, arising from the perturbation on the
phase mode due to the amplitude-phase coupling.

The results can
be tested experimentally.

\section*{Acknowledgements}
\noindent DSL would like to thank the Blackett Laboratory, Imperial
College for hospitality, where much of this work was performed and RR would like to thank the
National Dong Hwa University, Hua-Lien, for support and
hospitality, where some of this work was performed. The
work of DSL and CYL was supported in part by the National Science
Council and the National Center for Theoretical Sciences, Taiwan.

\end{document}